# Investigation of Al-Si-Cu alloys as phase change materials for high temperature thermal energy storage

Laura Teodorescu[1], Ángel Serrano[2], Kyran Williamson[2], Cristina Luengo[2], Artem Nikulin[2], Elena Palomo Del Barrio[2,3*], Grégory Largiller[1*]

[1] Université Grenoble Alpes, CEA, LITEN, DTCH, LCST, Avenue des Martyrs 17, Grenoble, 38000, France
[2] Centre for Cooperative Research on Alternative Energies (CIC energiGUNE), Basque Research and Technology Alliance (BRTA), Alava Technology Park, Albert Einstein 48 Vitoria-Gasteiz, 01510, Spain
3 Ikerbasque, Basque Foundation for Science, Bilbao 348013, Spain
[*]Corresponding authors: gregory.largiller@cea.fr, epalomo@cicenergigune.com

**Abstract**: The present work explores the suitability of Al-Cu-Si ternary alloys as high-temperature metallic phase change materials (PCMs) for thermal energy storage systems (TESS) operating between 550ºC and 850ºC. While prior research has primarily focused on thermodynamic modeling or thermal property characterization below 600°C, this work provides a comprehensive experimental assessment of selected invariant compositions within the Al-Cu-Si system. CALPHAD calculations were performed using FactSage 8.2 software (SGTE and FTLite database) to guide alloy selection, followed by synthesis and detailed characterization of melting point, latent heat, specific heat, thermal diffusivity, and thermal conductivity. Critically, the thermal reliability of these materials was evaluated under repeated solid-liquid cycling up to 100 cycles in oxidizing atmospheres, revealing their stability and degradation profiles. Additionally, dilatometry and density analysis were conducted to provide an in-depth understanding of the alloys and practical properties for end users. Among the tested alloys, several demonstrated high volumetric energy densities (over 500 kWh/m³ for a temperature difference of 300°C) and good thermal durability, establishing Al-Cu-Si alloys as promising PCM candidates for industrial-scale high-temperature energy storage applications. This study fills a notable gap in the literature by combining phase selection, comprehensive thermophysical property evaluation, and long-term cycling analysis above 600°C.

## 1. Introduction

In the last few decades, high-temperature thermal energy storage systems (TESS) have drawn a lot of interest due to their significance for renewable energy and their ability to supply industry with continuous high temperature (HT) heat. Based on these advantages, TES have been implemented in different fields, such as concentrated power plants (CSP), industrial waste heat recovery and peak regulating energy network [1]. Recent research highlight several established and emerging applications for TES such as thermal management in electronic devices, batteries and PV, thermal comfort in building, district heating, transportation as well in high-temperature industrial processes (temperatures over 1100°C) [2], ultra-high temperature storage for solar thermal generation, power-to-heat-to-power storage for district heating/electricity supply [3,4].

Nowadays, sensible heat storage in liquids or solids (e.g., oils, molten salts, rocks, ceramics, concrete) dominate the technologies due to their low cost. However, the volumetric energy density of sensible heat storage systems is quite low, resulting in a limitation for their use in applications where the space is at premium. Latent heat storage (LHS) based on phase change materials (PCM) is a very efficient method to significantly increase the volumetric energy density (x 5-10) for application where the available space is constrained [5–9].

Most high-temperature LHS systems rely on melting-solidification processes of anhydrous salts and metal alloys because of their high thermal stability. The advantage of the anhydrous salts and their mixtures compared to metal alloys is their relatively low-cost, whereas the metal alloys have much higher thermal conductivity, resulting more attractive in high-power, and relatively low vapor pressure applications [10,11].

For metallic PCMs (MPCM), pure metals come first to consideration, then simple eutectic alloys which allow to access intermediate melting temperatures [12]. As the optimal material's melting point is based on the operational temperature of the TESS, the MPCMs can be classified according to their melting temperature [3,13], their range being very wide, above 0°C, starting with 29.8°C for Ga up to 3420°C for W [14,15]. Besides the importance of the melting temperature of the MPCMs, comes the volumetric latent heat (ranging from 8.2Wh.m$^{-3}$ (Cs) to 1062.5kWh/m$^{-3}$ (Mo)) and the thermal conductivity which account for the storable amount of thermal energy and the time to release it respectively (from 13 W.m$^{-1}$.K$^{-1}$ (Bi) to 350.m$^{-1}$.K$^{-1}$ (Cu)) [5,14,16,17].

Based on recent studies MPCMs dedicated to high temperature applications can be divided into the following section:

- Zn-based and Mg-based alloys for the temperature ranges 300-450°C, mentioned in TES for industrial waste heat recovery [6,11].
- Al-based alloys for 450°C-660°C with some focused on CSP [11,18].
- Cu-based alloys for 660°C-850°, mostly on high-temperature thermal energy storage as advanced solar power systems [18]
- Fe-based alloys and Si-B alloys for temperatures above 900°C, focused on ultra-high temperature thermal energy storage for advanced concentrated solar power [19]

For high temperature TES system, aluminium eutectic alloys have been widely investigated as PCMs due to their high heat of fusion and their good thermal reliability. The focus was mostly on the eutectic point AlSi12, fist mentioned by Birchenal and Reichman [12]. The eutectic alloy has been further studied as heat storage in medium domestic heaters, steam generators, in heaters for electric and hybrid vehicles, heat exchanger for industrial heat waste recovery, in CSP and packed bed latent heat thermal energy storage (LHTES) systems [6,20–24].

In the search for PCMs with higher melting temperatures than the eutectic Al-Si alloys, Kotze et al. [7] identified pure aluminium as a promising PCM candidate. In Table 1 are presented the main thermophysical properties of pure Al and 88Al-Si12.

*Table 1 Thermophysical properties of Al and 88Al-12Si alloy at 20°C [7,9].*

| **Material** | **Melting Temp.** (°C) | **Latent Heat** ($J.g^{-1}$) | **Density** ($g.cm^{-3}$) | **Thermal Conductivity** ($W.m^{-1}.K^{-1}$) | **Volume Expansion** ($10^{-6}$ $K^{-1}$) |
|---|---|---|---|---|---|
| **Aluminium (Commercial purity)** | 660 | 395 | 2.7 | 273 | 80.1 |
| **88Al – 12Si** | 576 | 560 | 2.7 | 160 | 63.9 |

Aiming for a higher melting temperature range (550°C-850°C) in order to achieve a cascading effect of the PCMs, several promising candidates were introduced based on Al-Si systems [6,7,25]. Hence, Cu ($T_m$=1083°C) was implemented as a key alloying element to elevate the melting temperature of Al-Si alloys in order to reach in the desired temperature range. The Al content in the Al-Cu-Si alloys is thus considered as a key factor to improve the corrosion resistance of the alloys [26] and to modulate the melting point of the alloys in the temperature range targeted.

Al-Cu-Si alloys have gained significant attention due to their superior corrosion resistance compared to Al-Si and Al-Cu alloys [27,28]. The ternary system became popular in automotive and aerospace applications, especially due to their corrosion resistance, strength and lightweight compared to Al-Cu and Al-Si alloy respectively [29]. Notably, Ponweiser and Richter [28] provided a comprehensive overview of the phase composition of Al-Cu-Si ternary alloys. The authors aimed to reveal and solve new compounds with the interest in having a complete understanding of the Al-Cu Si system, being a key for further alloys developments.

Recent studies have been focused on Al-Cu-Si alloys as phase change materials for thermal storage applications. Rawson et al [30] focused on the ternary eutectic alloy Al - 25%Cu - 16%Si (%wt), within the temperature range 508 – 548 °C, having an enthalpy of 381.26 $J.g^{-1}$. Based on the study it was concluded that the alloy appeared to be a very good candidate as a phase change material used in the automobile industry, specifically for providing heat in electric vehicles.

Furthermore, Zhao et al. [31,32] studied the effect of Cu and the oxidation properties of the Al-Cu-Si alloys dedicated for latent heat energy storage (LTES). The MPCM candidates were selected within the temperature range of 600°C, with different Cu and Si contents (35-55% Cu content and 4.6%Si). The study revealed that the ternary alloys proved good thermal stability after high temperature oxidation. From an energetic point of view, Hallstedt et al [33] determined and compared with previous studies different invariant reactions in the Cu-Si and Al-Cu-Si system, along with their enthalpies. The invariant points covered a wide range of temperatures from 521°C up to 856°C.

The present study aims to develop and characterize several invariant points selected from the Al-Cu-Si ternary phase diagram as potential candidates for metallic phase change materials dedicated to high-temperature thermal energy storage systems (TESS) in industrial waste heat recovery. This research seeks to provide a comprehensive understanding of Al-Cu-Si alloys as suitable candidates for large-scale installation, addressing a gap in the literature, as no previous studies have been dedicated to this area. For this, preliminary assessments were made by CALPHAD approach using FactSage Thermochemical Software [34]. This made possible to define the points of interest and to calculate their expected energy, taking into account the possible discrepancies between databases. Once the selected candidates were synthetized, their microstructure and thermophysical properties (melting temperature, latent heat, thermal diffusivity and thermal conductivity) were characterized in order to provide a complete description of the developed alloys.

## 2. Material and Methods

### *PCM Preparation*

Granules of Al (purity 99,999%), Si (purity 99,999%) and Cu (purity 99,999%) were ordered from HMW Hauner GmbH & Co. KG. All the alloys were synthetized by high-frequency induction furnace with Cu coils linked to a high-frequency converter, using a cold copper crucible under primary vacuum. In addition, each sample was melted two to three times and turned upside down to guarantee high homogeneity. The studied alloys were synthetized by melting 100g for each sample. The synthesis parameters (input power, current, and frequency) were initially optimized for the 84Cu- 16Si %wt alloy. These parameters served as a reference for the synthesis of the rest of the alloys. The detailed parameters for the 84Cu- 16Si %wt sample are presented in *Table 2*.

*Table 2 Parameters used for 84Cu-16Si % wt sample for 100g weight.*

| **Direct Current Power** (kW) | **High-Frequency Voltage** (V) | **Direct Current** (A) | **Frequency** (kHz) | **Power Utilization Percentage** (%) |
|---|---|---|---|---|
| 6.4 | 215 | 35.9 | 130 | 12.0 |

### *Material Characterization*

SEM imagery (LÉO 1530 VP Gemini, Elektronenmikroskopie GmbH) on polished samples was used to observe the microstructure of synthetized alloys. The acquisition of spectra (EDX) was made by using an acceleration voltage from 5kV until 16kV. The standard corrections were calculated using the Esprit 1.9.4 software internal standard. The quantification of the chemical composition was obtained from an average of two areas of 0.11 $mm^2$ each.

The latent heat was measured using a STA 449 F3 JUPITER (Netzsch), used in TG-DSC mode, were each sample was placed in alumina crucible (85 µL), heated at 10°C.$min^{-1}$ and thermally cycled three times between 500°C and 900°C under $N_2$ flow (60mL.$min^{-1}$) to prevent oxidation.

For the thermal cycling of the selected MPCMs the study was performed under 10 $mL.min^{-1}$ of synthetic air, in order to better observe the corrosion phenomena.

The coefficient of linear thermal expansion (CLTE) was recorded using the thermo - mechanical analysis machine often also named dilatometry (SETSYS EVO SETARAM) on the temperature range adapted for each sample, with a rate of 2°$C.min^{-1}$, under Ar flow rate of 60 $mL.min^{-1}$. The mean coefficient of thermal expansion (CTE) was calculated from the following Equation (1):

$$\alpha = \frac{1}{T-T_0}\frac{D_T - D_{T0}}{e_0} \qquad (1)$$

where α ($10^{-6}$ °$C^{-1}$) is the mean coefficient of thermal expansion, $e_0$ length of the initial sample in mm, $D_{T0}$ and $D_T$ displacement from initial point and final point, $T_0$ and $T$ respectively initial and final temperature in the selected temperature range.

In addition, the apparent density was measured by Archimedes' Principle in water and ethanol at room temperature (20°C). To verify the accuracy of the measurements, the density ($\rho_0$) of Al and 88Al-12Si samples were measured to the value of 2.7 $g.cm^{-3}$ which is in agreement with literature [20,35].

The thermal diffusivity was determined by using the laser flash diffusivity method (LFA) on LFA-457 from NETZSCH. Samples with a side length of 10 mm and a thickness of 2 mm were used. In order to improve and promote the absorption and emission of heat, a graphite sprayed coating was added over the samples and the reference surfaces. Measurements were carried out under 50$mL.min^{-1}$ of dynamic Ar atmosphere. Furthermore, the thermal diffusivity was calculated using the "Parker equation" (2).

$$a = 0{,}1388\frac{L^2}{t1/2} \quad (3)$$

where α ($mm^2.s^{-1}$) is the thermal diffusivity of the sample, L (mm) is the thickness of the sample and t0.5 (s) is the time at which the rear surface temperature reaches 50% of its maximum increase. Thermal diffusivity measurements were conducted from room temperature up to 500°C, with three laser shots programmed at each temperature to ensure reproducibility. The Cape-Lehman model was selected for all alloys, as it proved to be the most suitable for the tested samples. To ensure accuracy, each shot was truncated once the temperature reached its maximum to exclude thermal losses from the calculations.

The specific heat capacity Cp of the studied materials was calculated using the LFA technique, with copper selected as the reference material. Copper (purity 99.999%) was chosen due to its well-known thermal properties and its similarity to the investigated samples under high-temperature conditions. In the LFA, Cp is determined by sequentially measuring the thermal response of both the reference material and the sample under identical conditions. The specific heat capacity of the sample is indirectly calculated using the following equation:

$$Cp^{sample} = \frac{T^{ref}}{T^{sample}}\frac{(\rho.l)^{ref}}{(\rho.l)^{sample}}\, Cp^{ref} \ (4)$$

Where, $Cp^{sample}$ and $Cp^{ref}$ are the specific heat capacities of the sample and reference material respectively. $\rho^{sample}$ and $\rho^{ref}$ are the densities of the sample and reference. $l^{sample}$ and $l^{ref}$, their thicknesses, and $\Delta T$ the temperature changes in the materials during the measurement. Once the thermal diffusivity data was obtained, Cp values were calculated by comparing the thermal responses of the samples to the copper reference. Measurements were performed under identical conditions using the Laser Flash system certified by NETZSCH.
The thermal conductivity ($\lambda$) was calculated by using the following equation:

$$\lambda (T)= \alpha(T)\ \rho (T)\ C_p(T) \qquad (2)$$

Where $\lambda$ ($W.m^{-1}.K^{-1}$) is the thermal conductivity at temperature T, $\alpha$ ($mm^2.s^{-1}$) is the thermal diffusivity at temperature T, $\rho$ ($g.cm^{-3}$) the density and Cp ($J.g^{-1}.K^{-1}$) the specific heat capacity at temperature T.

## 3. Results and Discussion

### *CALPHAD Calculations*

CALPHAD calculations were conducted using FactSage Thermochemical Software using FTLite or SGTE databases for alloys and FACTPS database for pure substance (all released in 2022). Ternary diagrams with liquidus projection were calculated to obtain ternary invariant points and find composition within the targeted melting range [34]. As databases were developed for a very large variety of alloys, none of them are dedicated to cover the whole Al- Cu-Si compositions over an extended range of temperature, thus some discrepancies between databases and with literature have been noticed.

The Figure 1 illustrates the phase diagram of Al-Cu-Si ternary system, which was prepared from "Phase Diagram" module using FTLite and SGTE 2022 database in FactSage 8.2 software, with the points of interest for the study.

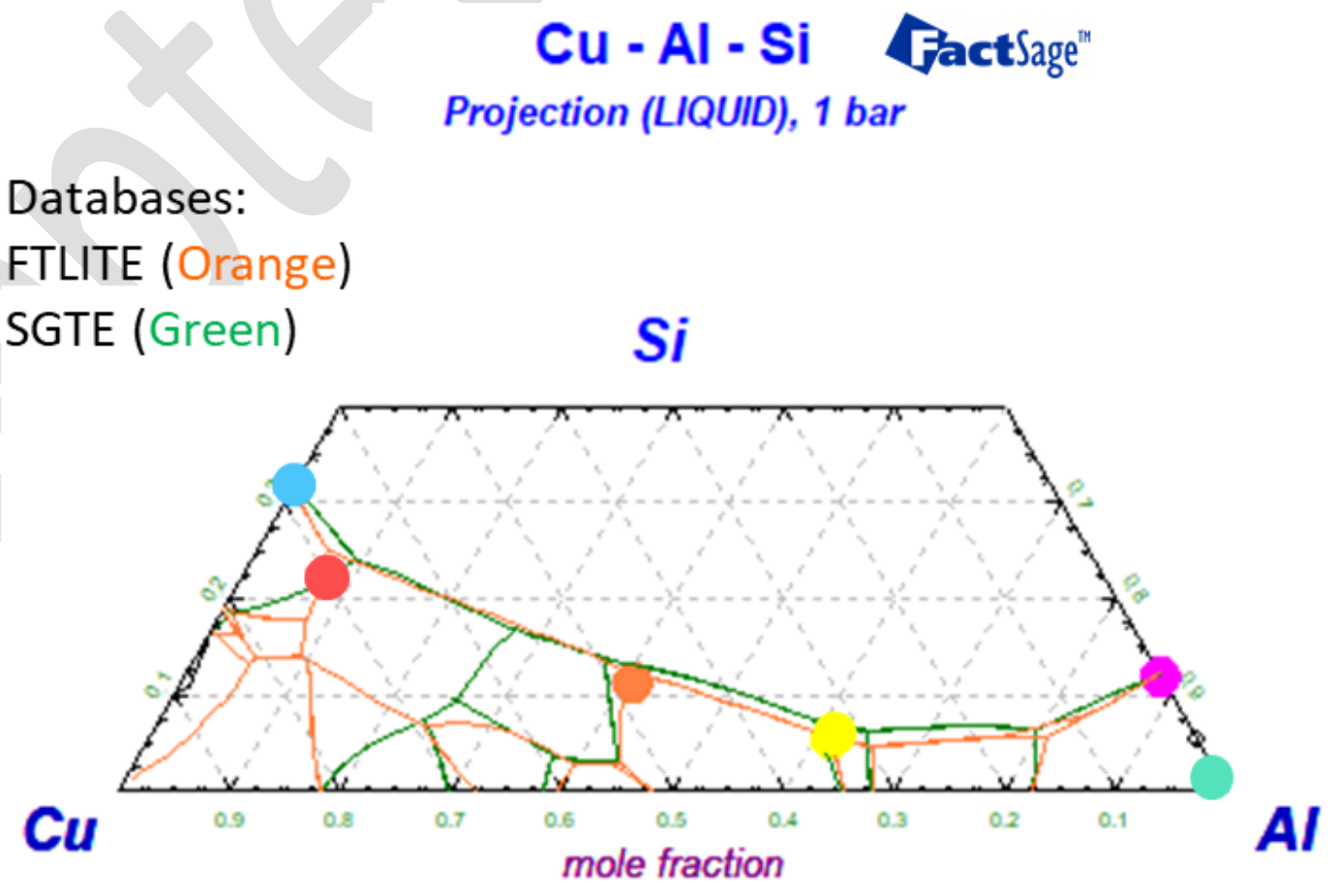

*Figure 1 Liquidus projection of ternary phase diagram Al-Cu-Si, using FTLITE and SGTE databases from FactSage 8.2 software.*

*Table 3* is summarizing the differences in latent heat and melting temperature for the selected alloys from the two databases. For pure Al databases gave the same values which are in agreement with literature [9]. In the case for 88Al-12Si (eutectic point), the melting point is at 578 °C with an enthalpy of 510 $J.g^{-1}$ using the SGTE database. However, with the database

FTLite it can be noted a slight difference such as the melting point at 577°C with the same heat of fusion 510 J.g$^{-1}$. Regarding the 84Cu-16Si alloy (eutectic point), SGTE database proposes a melting temperature 801°C associated to a heat of fusion 252 J.g$^{-1}$ (higher than FTLite). Nevertheless, the differences are small and still in agreement with the data from literature [6,11].

*Table 3 Comparison between databases FTLITE and SGTE for Al-Cu-Si alloys, focus on ΔH (latent heat) and $T_m$ (melting temperature)*

| ***PCMs*** | ***FTLite*** | | ***SGTE*** | |
|---|---|---|---|---|
| | ΔH ($J.g^{-1}$) | $T_m$ (°C) | ΔH ($J.g^{-1}$) | $T_m$ (°C) |
| **Al** | 404 | 660 | 404 | 660 |
| **88Al-12Si** | 510 | 578 | 510 | 577 |
| **84Cu-16Si** | 222 | 809 | 252 | 801 |
| **24.8Al-68.3Cu-6.9Si** | 462 | 725 | 462 | 690 |
| **4.1Al-84.2Cu-11.7Si** | 230 | 747 | 237 | 724 |
| **42.5Al-53.6Cu-3.9Si** | 438 | 577 | 437 | 558 |

Both 24.8Al-68.3Cu-6.9Si (peritectic point) and 4.1Al-84.2Cu-11.7Si alloys (eutectic point) have been presented in literature, by Hallstedt et al. [33], mostly from a thermodynamic point of view. For sample 4.1Al-84.2Cu-11.7Si using the SGTE Database, the transition to a liquid phase starts at 724°C with a latent heat of 237 J.g$^{-1}$. On the other hand, using FTLite database, the phase transition starts at a higher temperature such as 747°C with a latent heat of 230 J.g$^{-1}$. Comparing to Hallstedt et al. [33], the alloy show a melting temperature of 730°C with a latent heat of 141 J.g$^{-1}$. 24.8Al-68.3Cu-6.9Si alloy exhibits with FTlite database a latent heat of 462 J.g$^{-1}$ with onset temperature ($T_{onset}$) of 725°C and with SGTE a latent heat of 462 J.g$^{-1}$ with $T_{onset}$ of 690°C. For the same alloy, Hallstedt recorded a value of 173 J.g$^{-1}$ at a melting point of 740°C.

To investigate a varied range of temperatures in the ternary system Al-Cu-Si, the 42.5Al- 53.6Cu-3.9Si alloy (eutectic point) has been selected based on the liquidus projection of the phase diagram. This alloy presented a good latent heat value of 438 J.g$^{-1}$ at a melting temperature 577 °C while using FTLite database, the latent heat of 437 J.g$^{-1}$ at a melting temperature of 558°C was obtained for SGTE database.

These discrepancies between the databases confirmed the interest of studying the Al-Cu-Si alloys for thermal energy storage applications by DSC (differential scanning calorimetry) and TMA (thermomechanical analysis).

***Scanning Electron Microscopy Analysis***

SEM-BSE images have been taken on polished thick sections of the selected alloys. Figure 2 reveals the micrographs, noticing the presence of Si-rich formations, which are homogeneously distributed in the samples 84Cu-16Si and 24.8Al-68.3Cu-6.9Si.

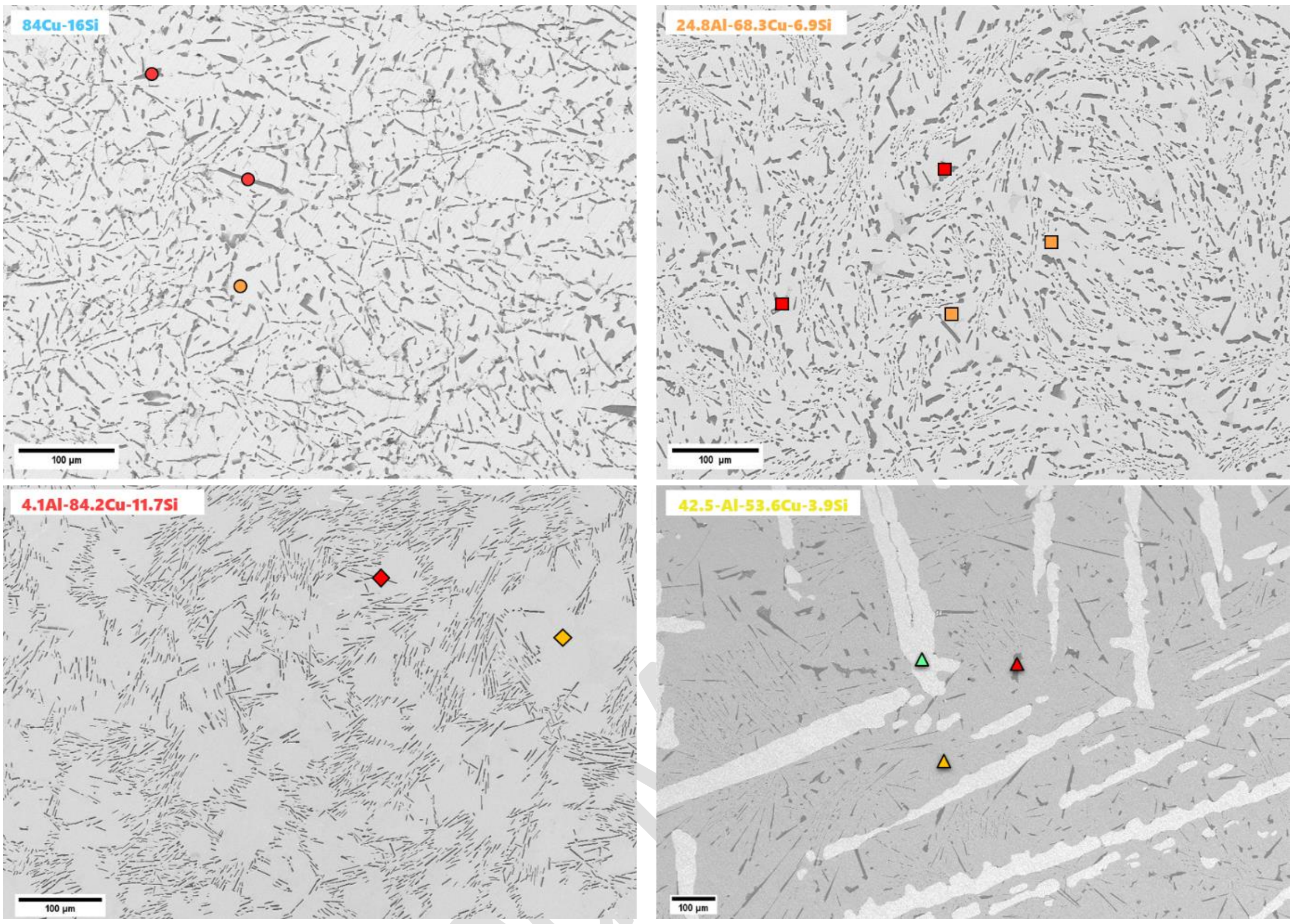

*Figure 2 SEM-BSE images acquired for samples 84Cu-16Si, 24.8Al-68.3Cu-6.9Si, 4.1Al- 84.2Cu-11.7Si and 42.5Al-53.6Cu-3.9Si, showing the presence of Si-rich formations in the matrix.*

SEM-EDX analysis was performed on bulk cross sections of the alloys, revealing their chemical composition, results shown in *Table 4.* A first step consisted in acquiring micro-images of four different compositions: 84Cu-16Si, 24.8Al-68.3Cu-6.9Si, 4.1Al-84.2Cu-11.7Si and 42.5Al-53.6Cu-3.9Si. The second step consisted in the quantification of the chemical elements of the MPCMs.

The 84Cu-16Si PCM alloy revealed a two-phase microstructure composed by Cu-rich matrix (bright area) and a dispersed Si-rich phase (dark needle shapes). The EDX point analysis confirmed this phase distribution. The Si-rich area (red dots) contained approximately 95.9 %wt. Si and only 2.8 %wt. Cu, confirming the presence of eutectic Si particles. Furthermore, the Cu-rich area (orange dots) highlighted the chemical quantification of 85 wt.% Cu and 11 %wt. Si, with a small oxygen contribution (~3.5 %wt.). The presence of minor oxygen signal likely comes from superficial oxidation during sample preparation.

The sample 24.8Al–68.3Cu–6.9Si %wt. highlighted a microstructure composed by Cu-Al matrix with dark elongated shapes. The punctual EDX analysis showed that the bright matrix (orange square) was mainly composed by Cu with 74 %wt., Al with 25 %wt. and O with 1% wt.

The dark needle shapes (red squares) revealed the Si-rich precipitates, as observed in the 84Cu-16Si alloy. As before, the minor oxygen signals are attributed to surface oxidation during polishing.

The SEM images acquired for 4.1Al-84.2Cu-11.7Si %wt. revealed a Cu-rich microstructure represented by a bright matrix and dark needle-like forms. The EDX analysis of the bright matrix region (orange diamond) showed a composition of Cu 87 %wt., Si with 10 %wt., Al with 1.26 %wt. and O with 2 %wt. with. The dark acicular regions (red diamond) showed a significant Si content consistent with Si-rich intermetallic**.** Overall, the main morphology indicated a two-phase microstructure with a Cu-rich matrix and Si-rich needles**.**

The last sample, 42.5Al–53.6Cu–3.9Si (wt.%) alloy highlighted a microstructure, mainly composed of an Al-rich matrix**,** bright Cu–Al intermetallic phases, and dark Si-rich needles**.** EDX analysis of the bright regions (green triangle) showed a composition of approximately 70%wt of Cu**,** 30%wt. Al and 1.6 %wt. of oxygen**,** which is consistent with Cu–Al intermetallic compounds**.** The grey matrix (orange triangle) contained 44 %wt. of Al**,** 53.4 %wt. Cu and 1 %wt. of Si. The dark needles shapes (red square), represented as before, are composed by Si-rich precipitates (98 %wt.)**.**

It should be pointed out that the emission ray Kα of the Al and Si are close in energy (1.486 keV for Al and 1.740 keV for Si). Therefore, a peak overlap and interference was noted, for the sample 24.8Al-68.3Cu-6.9Si, which lead to a little mismatch of the quantified elements and the original composition. EDX demonstrates the presence of oxygen on the prepared cross section. As mentioning above this oxygen can merely be attributed to the presence of oxides on the surface of the SEM samples during polishing. Some further investigation based on X-Ray diffraction demonstrated the formation of CuO in 84Cu-16Si, $Al_2O_3$ (theta and corundum) on Al, $SiO_2$ in 88Al-12Si and a mixture of $SiO_2$ and CuO on Al-Cu-Si alloys.

*Table 4 Chemical composition (%wt), obtained by SEM-EDX on thick sections (±standard deviation)*

| PCM | Element | | | |
|---|---|---|---|---|
| | Al | Cu | Si | O |
| 84Cu-16Si | - | 74.4 ±0.3 | 20.0 ±0.3 | 5.6 ±0.1 |
| 24.8Al-68.3Cu-6.9Si | 19.7 ±0.6 | 68.2 ±0.5 | 11.2 ±0.1 | 0.8 ±0.1 |
| 4.1Al-84.2Cu-11.7Si | 5.5 ±1.4 | 81.2 ±1.1 | 14.1 ±0.1 | 0.8 ±0.7 |
| 42.5Al-53.6Cu-3.9Si | 39.3 ±0.1 | 52.3 ±0.2 | 6.6 ±0.2 | 1.8 ±0.1 |

***Density Measurements***

The density measures made it possible to calculate the total thermal energy that can be stored by a selected PCM per volume. Table 5 shows the recorded density values for each alloy, using the Principle of Archimedes. Two tests have been performed, one in water and another in ethanol. As expected, the addition of Cu increases the density of the MPCMs.

*Table 5 Density results obtained by Principle of Archimedes, measurements performed in water and ethanol solution (± standard deviation)*

| PCM | Density (g.cm$^{-3}$) | |
|---|---|---|
| | Water | Ethanol |
| Al | 2.68 ±0.03 | 2.70 ±0.01 |
| 88Al-12Si | 2.63 ±0.01 | 2.63 ±0.01 |
| 84Cu-16Si | 6.78 ±0.01 | 6.79 ±0.01 |
| 24.8Al-68.3Cu-6.9Si | 5.09 ±0.01 | 5.09 ±0.01 |
| 4.1Al-84.2Cu-11.7Si | 6.86 ±0.01 | 6.88 ±0.01 |
| 42.3Al-53.6Cu-3.9Si | 4.33 ±0.01 | 4.31 ±0.01 |

### *Differential Scanning Calorimetry*

For every synthetized alloy, DSC was performed with three successive thermal cycles. The thermographs obtained for each alloy are depicted on Figure 3 with the 3$^{rd}$ heating ramp and compared with the FactSage calculated energies from the two databases. The results show that through the refinement of compositions and synthesis processes, MPCMs exhibit phase transitions across the targeted temperature range of 550°C to 850°C. Notably, these materials demonstrate phase transition enthalpy values exceeding 180 J.g$^{-1}$, with the 88Al-12Si alloy achieving values surpassing 450 J.g$^{-1}$. Furthermore, the experimental results show good alignment with thermodynamic calculations derived from CALPHAD-based computations. However, certain discrepancies were observed, particularly in the ternary alloy systems, such as Al-Cu-Si, when comparing the outputs from different databases like FTLite and SGTE.

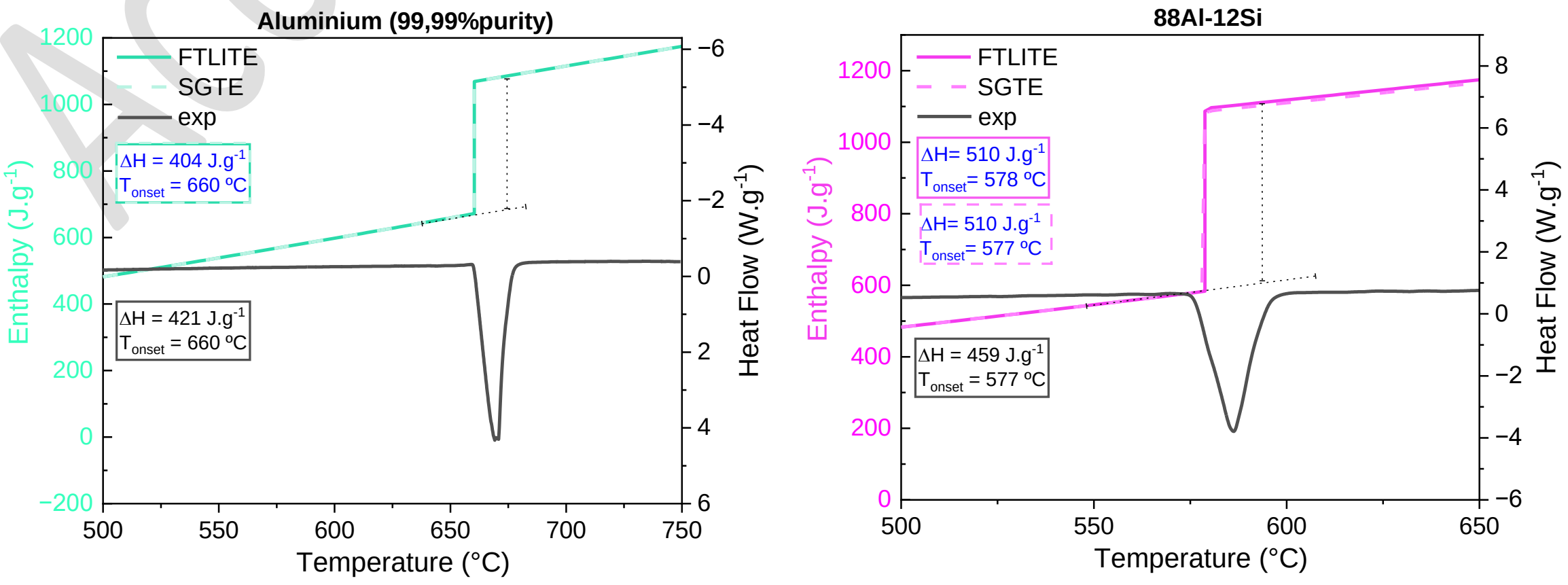

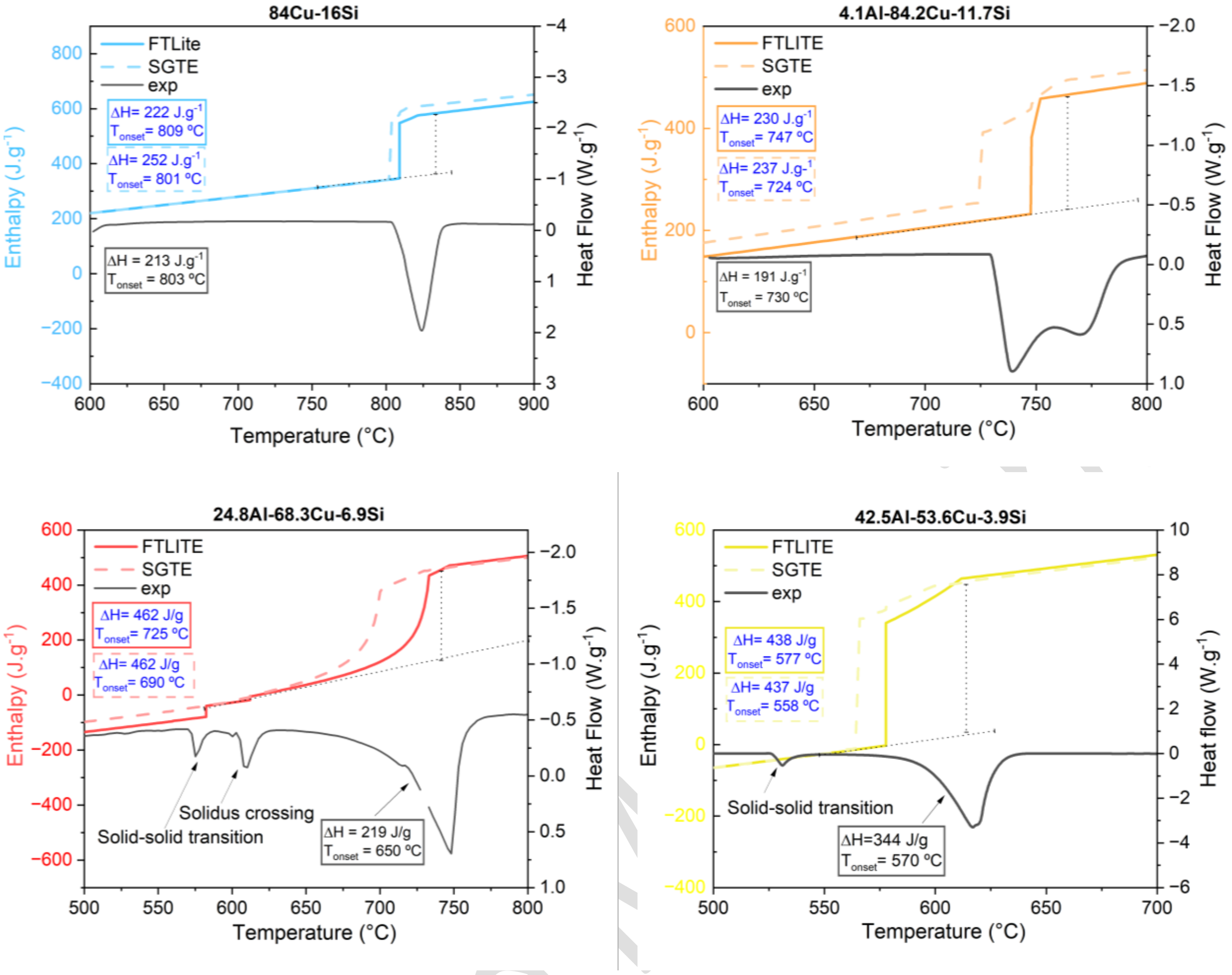

*Figure 3 Thermographs of DSC experimental results compared to CALPHAD calculations for FTLite and SGTE databases: for Al, 88Al-12Si, 84Cu-16Si and Al-Cu-Si alloys*

Comparing the obtained results in the actual research with Hallstedt et al. [32] thermochemical study in the Al-Cu-Si system, several differences can be noted (Table 6). Starting with the sample 84Cu-16Si, the measurement by DSC recorded a melting temperature of 803°C during the third cycling with a latent heat of 213 $J.g^{-1}$ as visible on the binary phase diagram on Figure 6. This revealed a difference of 22% lower than the literature value of 271.8 $J.g^{-1}$. In contrast with the FactSage simulation, FTLite database is 4% higher while SGTE database is 18% higher than the experimental latent heat.

Sample 24.8Al-68.3Cu-6.9Si experiences a progressive melting behaviour from 625°C with a latent heat of 219 $J.g^{-1}$. This value is more than twice lower than the one calculated by Calphad method. Both FTLite and the experimental curve show a solid-to-solid phase transformation around 575°C which is not taken into account by the SGTE database. The later transition recorded at 600 °C was found to represent the solidus line crossing. The sample 4.1Al-84.2Cu-11.7Si showed a melting temperature of 730°C with a latent heat of 191 $J.g^{-1}$, which is 74% higher than the literature value of 110 $J.g^{-1}$. The comparison with FactSage calculations highlighted a discrepancy reaching 23 % higher for the FTLite and a bit less for the SGTE database. Regarding the shape of the simulated curves during melting, SGTE database demonstrates the progressive dissolution of the FCC into the liquid phase observed on the experimental DSC.

The last sample, 42.5Al-53.6Cu-3.9Si, the DSC latent heat measures was 344 $J.g^{-1}$ with the temperature onset of 570 °C. Comparing the results with the FactSage calculations, the latent heat is 27% higher for both databases. The calculated transition was supposed to occur at non invariant reaction, this mechanism was not clearly underlined by the experiment.

*Table 6 Comparison between laboratory results, Literature from *Shamberger et al. □Hallstedt et al.* [17,33] *and FactSage Calculations*

| **PCM** | **DSC (under $N_2$)** | | **Literature*□** | | **FactSage** | | | |
|---|---|---|---|---|---|---|---|---|
| | III^rd cycle | | | | FTLite | | SGTE | |
| | TEMP Onset (°C) | ENDO ($J.g^{-1}$) | TEMP On set (°C) | DSC ($J.g^{-1}$) | TEMP Onset (°C) | ENDO ($J.g^{-1}$) | TEMP Onset (°C) | ENDO ($J.g^{-1}$) |
| **Al** | 660 | 421 | 660.3* | 397* | 660 | 404 | 660 | 404 |
| **88Al-12Si** | 577 | 459 | 579* | 498* | 578 | 510 | 577 | 510 |
| **84Cu-16Si** | 803 | 213 | 803□ | 271.8□ | 809 | 222 | 801 | 252 |
| **24.8Al-68.3Cu-6.9Si** | 650 | 219 | 740□ | 173□ | 725 | 462 | 690 | 462 |
| **4.1Al-84.2Cu-11.7Si** | 730 | 191 | 730□ | 110□ | 747 | 230 | 724 | 237 |
| **42.5Al-53.6Cu-3.9Si** | 570 | 344 | - | - | 577 | 438 | 558 | 437 |

In summary, the DSC measurements for all samples showed significant differences when compared to both literature values and FactSage calculations. These discrepancies highlight the need for further investigation into the thermodynamic properties of these PCMs and the accuracy of the existing databases even if the proposed Calphad tool remains relevant for the study. These variations are expected in intermediate compositional regions of ternary phase diagrams, where the precision of database inputs is less consistent. This limitation highlights the importance of validating computational results with experimental data and literature review, especially for more complex alloy systems. However, the global agreement consistency underscores the interest of computational models in guiding material selection and optimization efforts.

In addition, Figure 4 presents the specific heat capacity values of the studied PCMs from room temperature to 500°C, where significant differences are observed between the PCMs, especially a possible phase transition in the binary Al-Si system between 150 °C and 200 °C.

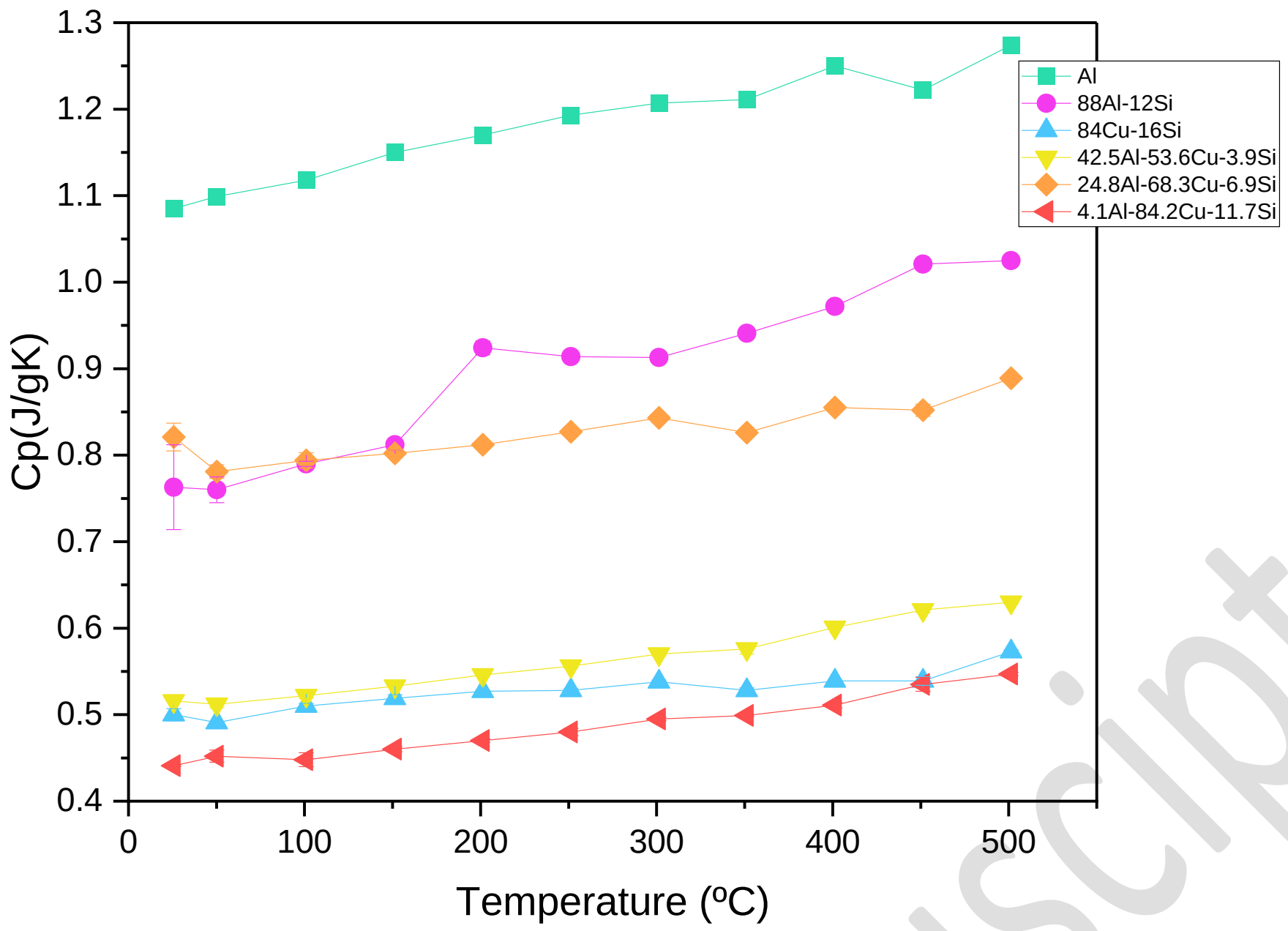

*Figure 4 Specific heat capacity values for studied PCMs.*

***Dilatometry Analysis***

Dilatometry is a very efficient technique to measure the coefficient of linear thermal expansion but also to study the phase transitions of the solid phase during heating as it enables to identify transitions below the detection capability of DSC. The CLTE of pure Al and 88Al- 12Si alloys were in accordance with previous authors work with measured values of 31.82 $10^{-6}$ °$K^{-1}$and 22.01 $10^{-6}$ °$K^{-1}$ respectively [20,35,36]. Al-Si system is composed of two phases: Si and Al rich phase α with volume proportion increase when heating. On Figure 5 the increase of the α phase in the alloy by Si addition reduces the CLTE of pure Al. Cu-Si system is more complex than Al-Si system. At 777 °C, pure Cu has a CLTE of 21.90 $10^{-6}$ °$K^{-1}$ [37], the recorded value of 84Cu-16Si was 21.05 $10^{-6}$ °$K^{-1}$ between 150 °C and 770 °C, all the other mean CLTE values are between Al and 84Cu-16Si which form the limits of possible mean values reported in *Table 7*.

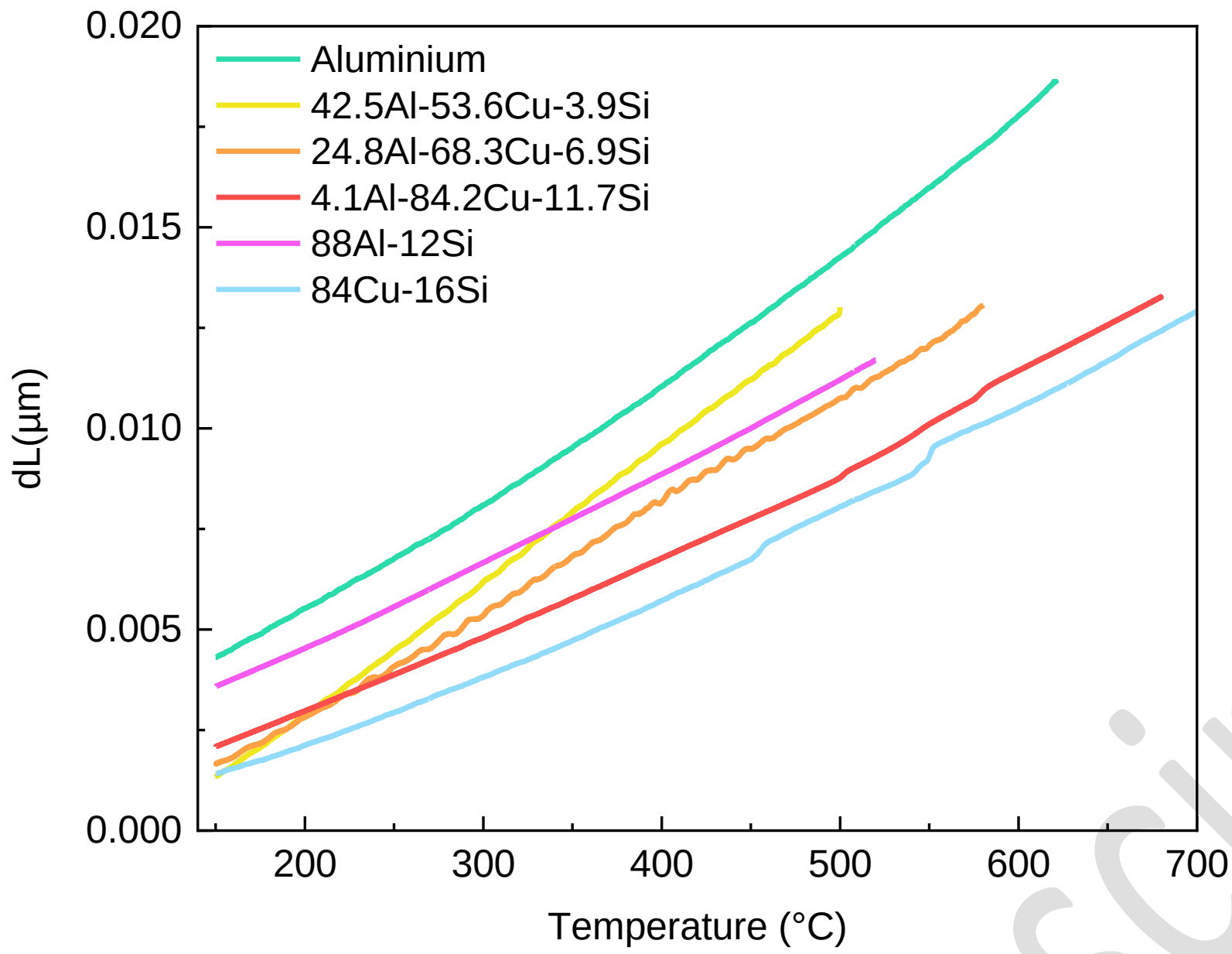

*Figure 5 Displacement dL (µm) of each selected PCM, selected PCM during second cycle*

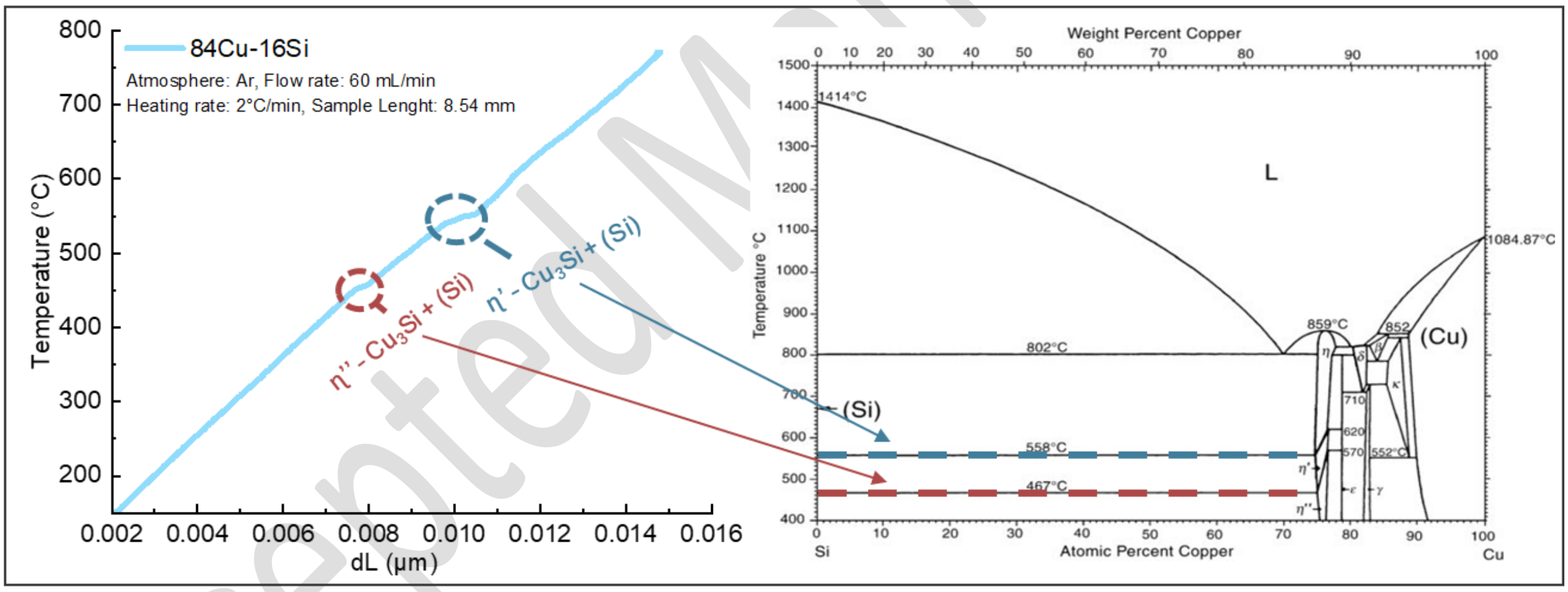

*Figure 6 Dilatometry measurement on 84Cu-16Si (left image) and phase diagram of Cu-Si according to H.Okamoto (2016) (right image)*

88Al-Si12 and 24.8Al-68.3Cu-6.9Si revealed a linear thermal expansion as 42.5Al- 53.6Cu-3.9Si, 4.1Al-84.2Cu-11.7Si and 84Cu-16Si have thermal artefacts which suggest solid to solid phase changes. With reference to copper binary (84Cu-16Si) in Figure 6 the presence of two phases recorded at 452°C and 538°C correspond to ɳ'' – Cu3Si + (Si) formation theoretically formed at 467°C and ɳ' – Cu3Si + (Si) formed at 558°C [33,38,39], these phases are not described in SGTE and FTLite databases.

*Table 7 Mean CLTE recorded for each alloy from 150°C to specified end temperature value. Data recorded from the second cycle.*

| PCMs | Start (°C) | End (°C) | Mean Alpha ($10^{-6}$/°C$^{-1}$) |
|---|---|---|---|
| Al | 150 | 620 | 31.82 |
| 88Al-12Si | 150 | 520 | 22.01 |
| 84Cu-16Si | 150 | 770 | 21.05 |
| 24.8Al-68.3Cu-6.9Si | 150 | 580 | 26.48 |
| 4.1Al-84.2Cu-11.7Si | 150 | 680 | 21.19 |
| 42.5Al-53.6Cu-3.9Si | 150 | 500 | 32.80 |

***Thermal cycling analysis***

In section Differential Scanning Calorimetry, PCMs were evaluated through 3 successive thermal cycles, under neutral atmosphere. A key point in the technological transfer is the evaluation of the thermal cycling performance of PCMs for long-term reliability of the TES systems, particularly in industrial applications where air is often used as heat transfer fluid. In this section, thermal cycling experiments are conducted over 100 cycles five PCMs materials to assess their stability. DSC measurements at cycles 0, 20, 40, and 100 provide insight into potential shifts in phase transition temperatures and enthalpy losses. The Table 8 show the values obtained by DSC over the 3rd cycles, including the enthalpy obtained during the phase transition during the heating and cooling steps. As can be observed, Al demonstrates exceptional stability throughout the thermal cycling process. The enthalpy losses, calculated from the endothermic process during heating and referenced to the first measurement obtained in air, confirm that Al retains more than 96% of its initial latent heat even after 80 cycles, making it a highly reliable material for applications requiring long-term durability under thermal cycling. Furthermore, 88Al-12Si retains approximately 85% of its initial latent heat, with the majority of the reduction occurring within the first 20 cycles. After this initial adjustment, the material exhibits stable performance across subsequent cycles. This behavior may be attributed to a passivation layer created due to the initial oxidation of aluminium to form a stable and protective alumina layer [39]. The alloy 4.1Al-84.2Cu-11.7Si shows a notable evolution of its phase transition enthalpy over the cycling process, with a final retention close to 88% of its initial latent heat. The progressive stabilization of the enthalpy curve suggests that the eutectic composition may not have been fully synthesized initially. The thermal cycling potentially facilitates further alloy homogenization or phase redistribution, resulting in an improved and stabilized performance over time. Additionally, we measured the mass evolution during cycling and no mass uptake was recorded indicating that high oxidation mechanisms might not be responsible of the enthalpy change. The binary alloy 84Cu-16Si exhibits significant enthalpy degradation, retaining ~63% of the initial enthalpy by 100 cycles. This decline suggests considerable structural or chemical instability under thermal cycling, likely due to phase reorganization (see SI). Last, the ternary alloy 24.8Al-68.3Cu-6.9Si presents an enthalpy curve of this alloy shows a continuous drop, retaining a 65% of its initial enthalpy by the end of the test.

*Table 8 Enthalpy values recorded for every PCM over multiple thermal cycles under air, with the pristine sample (3rd cycle of DSC, under $N_2$), during the endothermic and exothermic phase transition.*

| | Cycles | Endothermic phase ($J.g^{-1}$) | Exothermic phase ($J.g^{-1}$) |
|---|---|---|---|
| Al | Pristine ($N_2$) | 421 | 438 |
| | 3 | 411 | 422 |
| | 20 | 400 | 409 |
| | 40 | 394 | 401 |
| | 80 | 398 | 411 |
| 88Al-12Si | Pristine ($N_2$) | 459 | 465 |
| | 3 | 533 | 544 |
| | 20 | 453 | 465 |
| | 40 | 448 | 464 |
| | 100 | 452 | 450 |
| 84Cu-16Si | Pristine ($N_2$) | 213 | 180 |
| | 3 | 260 | 240 |
| | 20 | 190 | 190 |
| | 40 | 159 | 168 |
| | 100 | 163 | 161 |
| 24.8Al-68.3Cu-6.9Si | Pristine ($N_2$) | 291 | 411 |
| | 3 | 318 | 361 |
| | 20 | 222 | 257 |
| | 40 | 211 | 254 |
| | 100 | 207 | 198 |
| 4.1Al-84.2Cu-11.7Si | Pristine ($N_2$) | 191 | 194 |
| | 3 | 212 | 212 |
| | 20 | 176 | 174 |
| | 40 | 211 | 214 |
| | 100 | 186 | 183 |

***Thermal Conductivity and Diffusivity***

Figure 7 illustrates the thermal conductivity values for the studied materials from room temperature to 500°C. The results were calculated by taking into account the Cp and thermal expansion measured, they indicate significant variations in thermal conductivity among the alloys. Pure aluminium exhibits the highest thermal conductivity. For the 88Al-12Si alloy, thermal conductivity is slightly reduced compared to pure aluminium, showing a similar decreasing trend. In contrast, alloys with higher concentrations of silicon and copper, such as 84Cu-16Si and Al-Cu-Si compositions, exhibit considerably lower thermal conductivity. These trends correlate well with the structural complexity and the influence of alloying elements that hinder heat transfer.

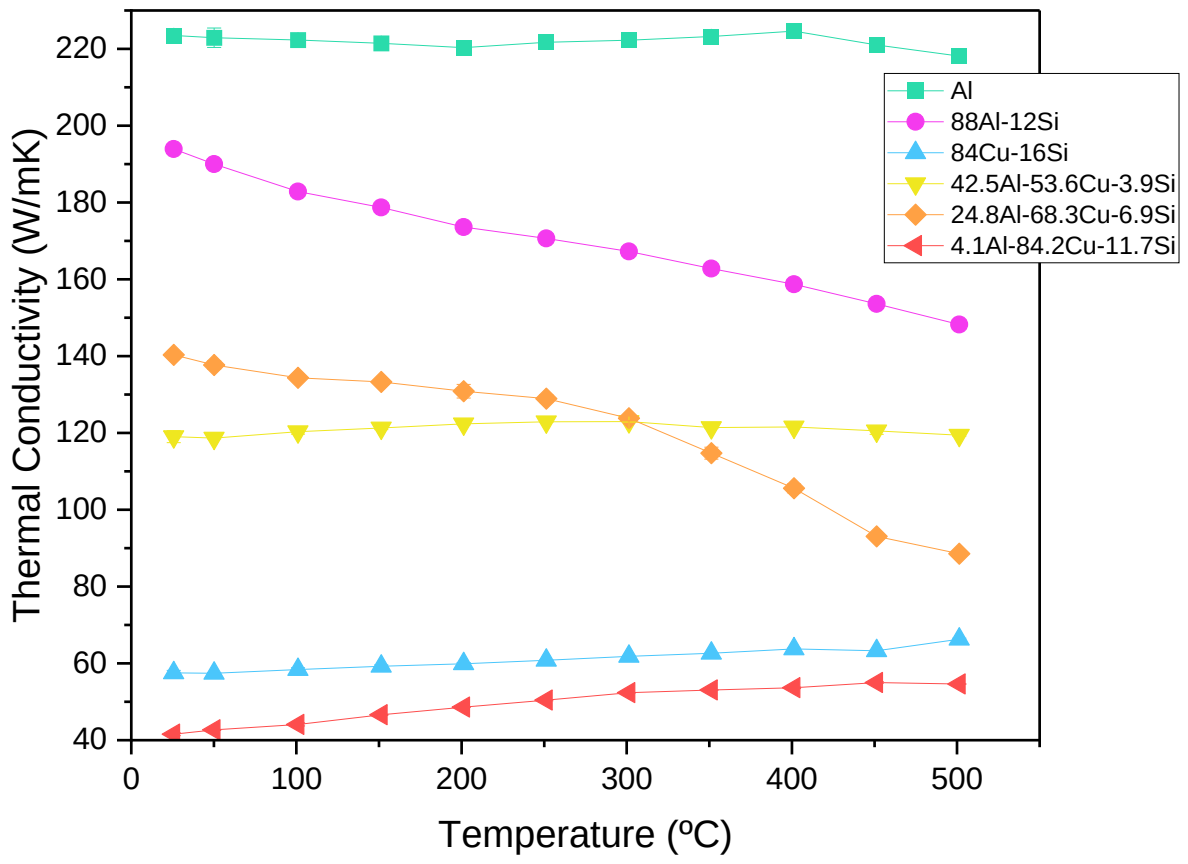

*Figure 7 Thermal conductivity values for the studied PCMs.*

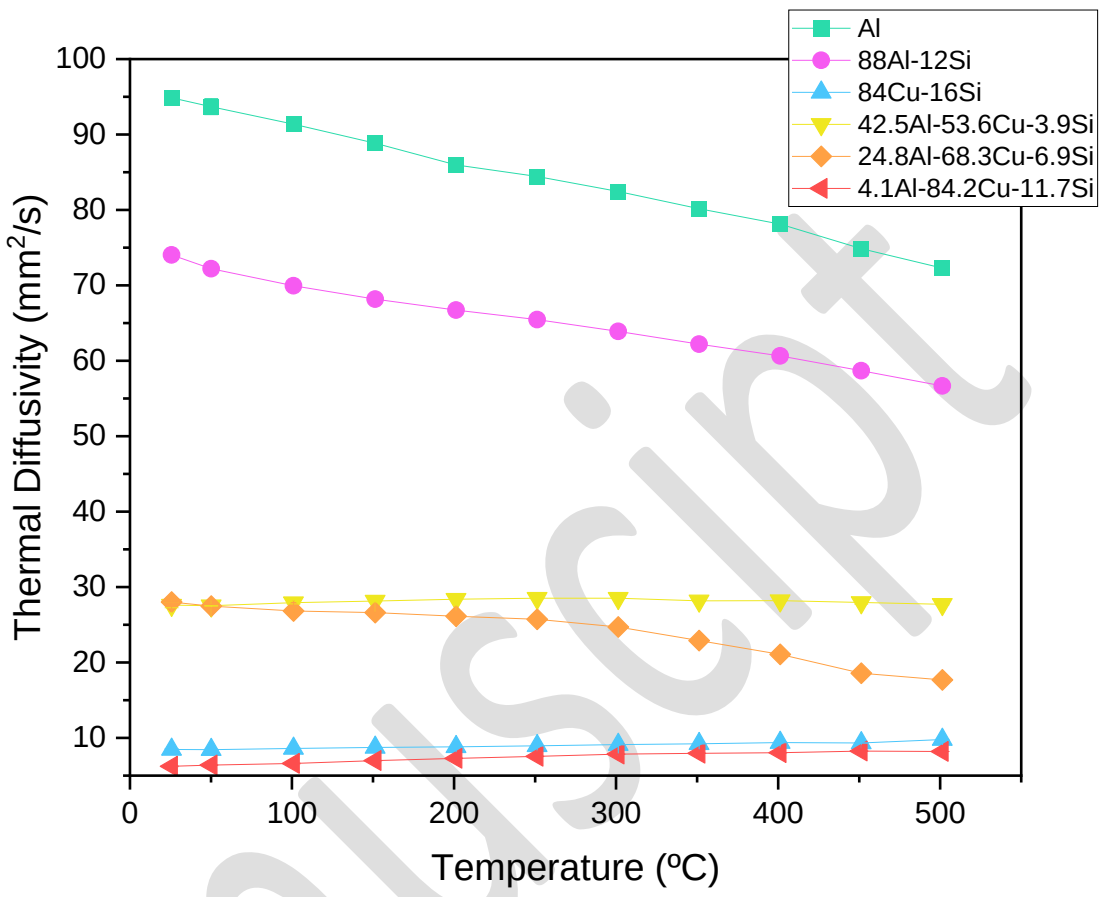

*Figure 8 Thermal diffusivity of each PCM over the studied temperature range.*

The thermal diffusivity values obtained for the studied alloys from room temperature to 500°C are presented in **Figure 8.** The results demonstrate significant differences in thermal diffusivity across the alloys. Pure Al exhibits the highest diffusivity, followed by 88Al-Si12, with values decreasing as the temperature increases, indicative of its high thermal conductivity and low heat capacity.

Al-Cu-Si compositions, show intermediate diffusivity values due to the combined influence of aluminium and the lower-conductivity alloying elements. 84Cu-16Si and Cu-rich compositions exhibit the lowest thermal diffusivity values.

Interestingly, the results for Al-Cu-based alloys with higher copper content, such as 42.5Al- 53.6Cu-3.9Si % wt and 24.8Al-68.3Cu-6.9Si %wt, deviate from the typical trend observed in the literature, where thermal conductivity usually decreases with temperature[41]. Instead, these alloys display a relatively stable or slightly increasing diffusivity with temperature. Similar behaviors have been reported in studies on Al-Cu alloys [42], which attribute such trends to microstructural phenomena such as grain refinement and precipitate coarsening at elevated temperatures. These mechanisms could increase the mean free path for heat carriers, partially offsetting the expected decrease in diffusivity.

To better understand the unexpected behavior in high-Cu Al-Cu alloys, future work should integrate detailed microstructural characterization (e.g., grain size, porosity, intermetallic phase analysis), compositional studies, and thermal testing. Correlating these findings with phase diagrams and computational thermodynamic predictions could shed light on the interplay between microstructure, composition, and thermal properties.

## 4. Conclusion

This research made possible to develop and investigate Al-Cu-Si PCMs dedicated to high temperature thermal storage systems for industrial applications between 550 °C and 850 °C. The results demonstrated that the studied Al-Cu-Si alloys can indeed contribute significantly in achieving high storage densities for industrial thermal energy storage systems. Even if at a first glance, Cu–Si alloys show excellent thermal storage capacity, their cycling stability is significantly reduced when tested in air. Additionally, the addition of Cu into Al-Si alloys doesn't improve their thermal conductivity, limiting the overall heat transfer efficiency. On the other hand, the Al and Al-Si alloy revealed superior thermal transport characteristics**,** which can enhance the performance of a purely sensible heat storage system.

*Table 9 Melting Temperature, Thermal Energy Storage Capacity (kWh/m$^3$) (including the sensible latent heat) in the temperature range 300°C, thermal conductivity for metal-based PCMs*

| **PCM** | **Melting T under $N_2$** (°C) | **TES Capacity** $\Delta T$ = 300°C (kWh/m$^3$) | **Thermal conductivity at 25°C** (W.m$^{-1}$.K$^{-1}$) |
|---|---|---|---|
| **Al** | 660 | 583 | 223 |
| **88Al-12Si** | 577 | 535 | 193 |
| **84Cu-16Si** | 803 | 705 | 57 |
| **24.8Al-68.3Cu-6.9Si** | 650 | 667 | 140 |
| **4.1Al-84.2Cu-11.7Si** | 730 | 647 | 41 |
| **42.5Al-53.6Cu-3.9Si** | 570 | 619 | 119 |

Therefore, for industrial application, thermal cycling tests under air revealed a distinct ranking in the stability of the materials. Based on the results we classify the stability of the PCMs as:
**Excellent Stability**: Pure aluminium (Al) is the best-performing material, retaining over 95% in both melting and solidification of its latent heat with no significant deterioration, making it an ideal candidate for high-durability systems.
**Good Stability**: 88Al-12Si and 4.1Al-84.2Cu-11.7Si show promising stability, with stable enthalpy retention of over 85% after 100 cycles. The eutectic alloy's evolution over cycles suggests potential optimization during use, which could further enhance its performance.
**Poor Stability**: 84Cu-16Si and 24.8Al-68.3Cu-6.9Si, exhibit significant degradation, particularly in the exothermic phase. These materials are unsuitable for long-term thermal storage applications in air environments.
For metal-based PCMs, scaling up the testing of Al, 88Al-12Si and the 4.1Al-84.2Cu-11.7Si alloy is crucial to validate their performance under operationally relevant conditions, particularly in terms of stability. These materials have shown promising stability and energy storage densities, making them strong candidates for large-scale deployment. Finally, it is critical to note that the selection of materials for scaling up will be significantly influenced by the outcomes of the PCM-refractory compatibility studies.

## Acknowledgements

This research is part of a large European project, HEATERNAL, which its main ambition is to meet the industry needs of a constant high temperature (HT) heat in the face of climate and geopolitical urgencies. This document has been prepared within the framework of IEA ES TCP. This project has received funding from the European Union's Horizon Europe research and innovation program under grant agreement No 101103921, HEATERNAL.

# Supplementary informations

## Two diagrams with liquidus projections (FTLite and SGTE)

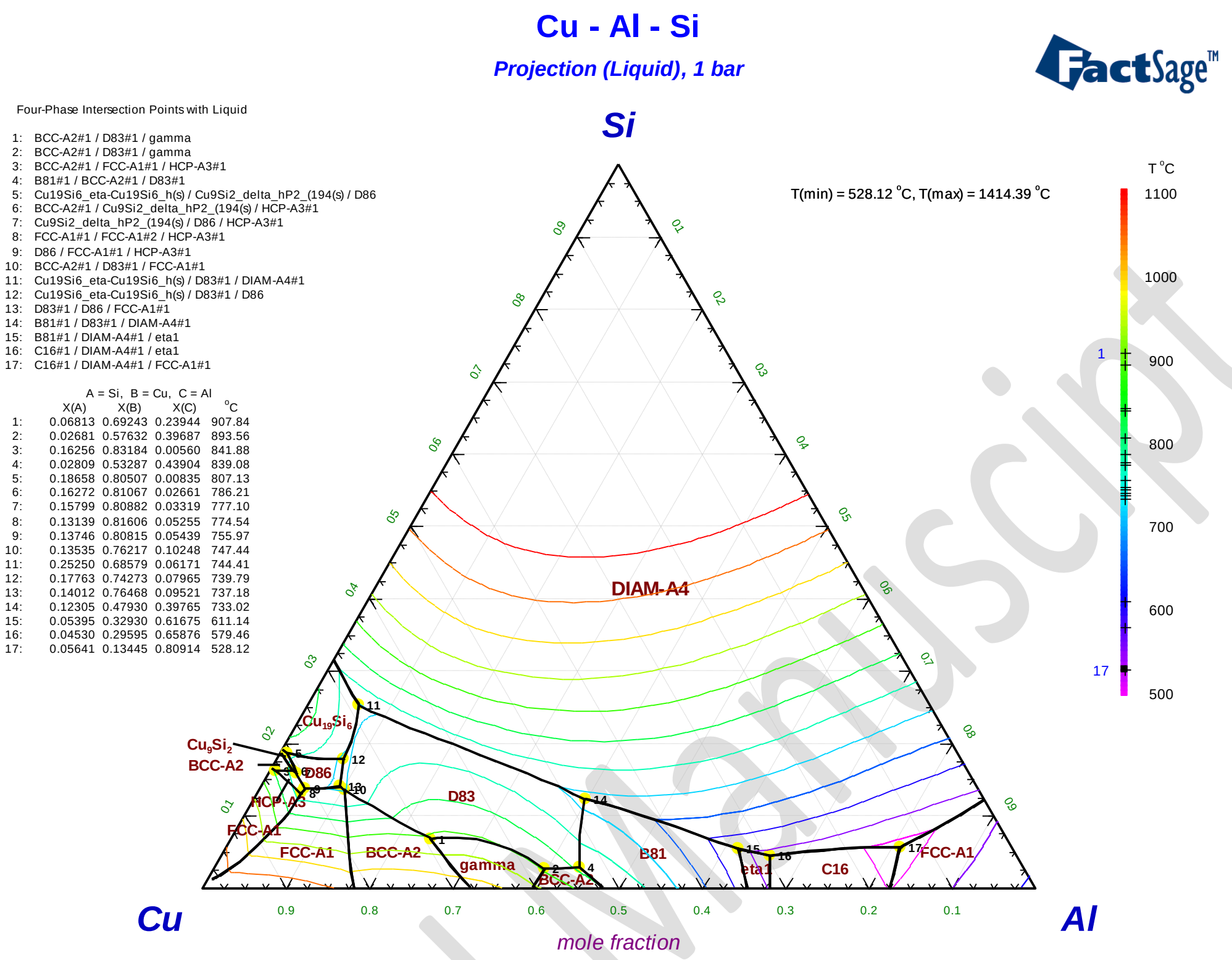

Figure 1-A Diagram with liquidus projections- FTLite

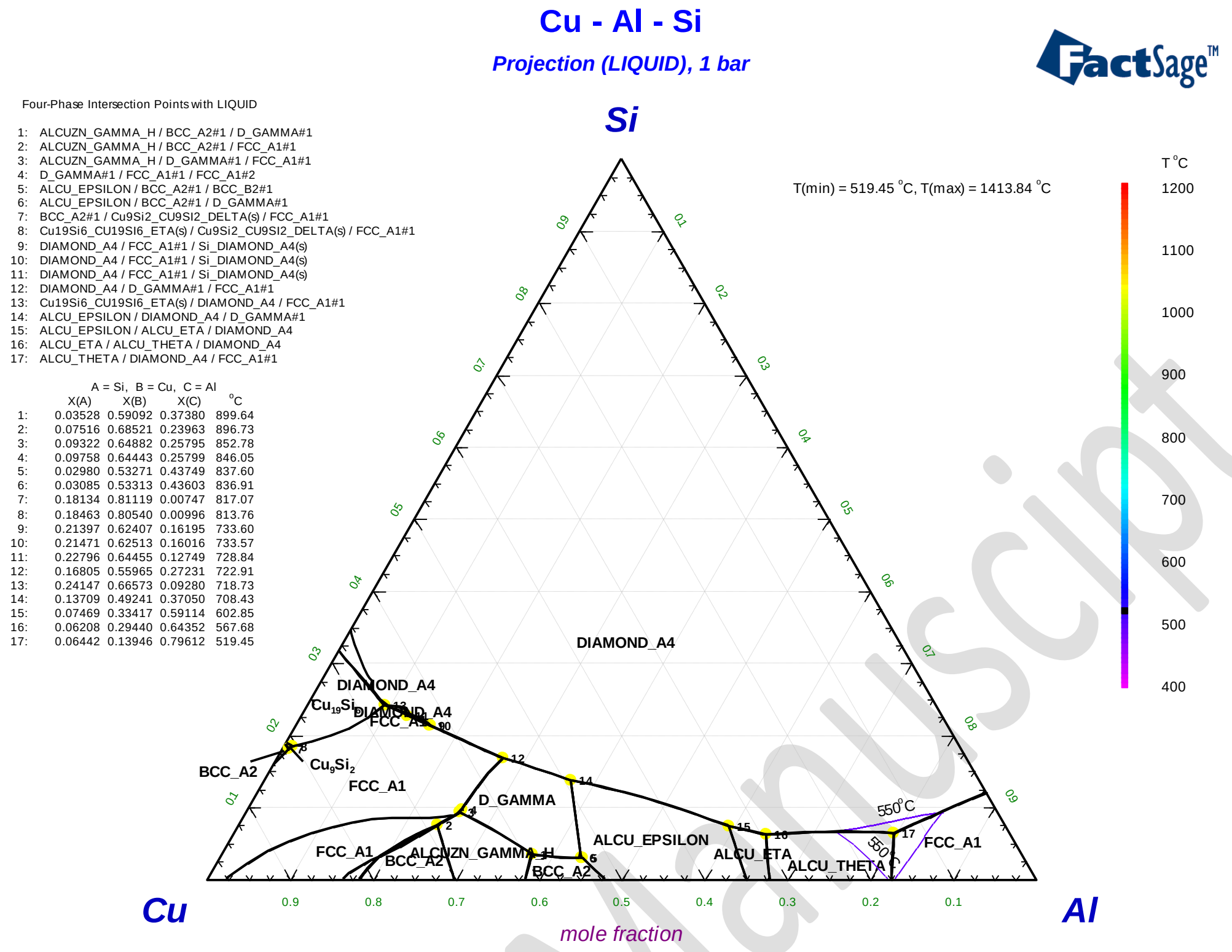

Figure 2-A Diagram with liquidus projections – SGTE without FCC0 phase (either FCC0 or FCC2 has to be selected)

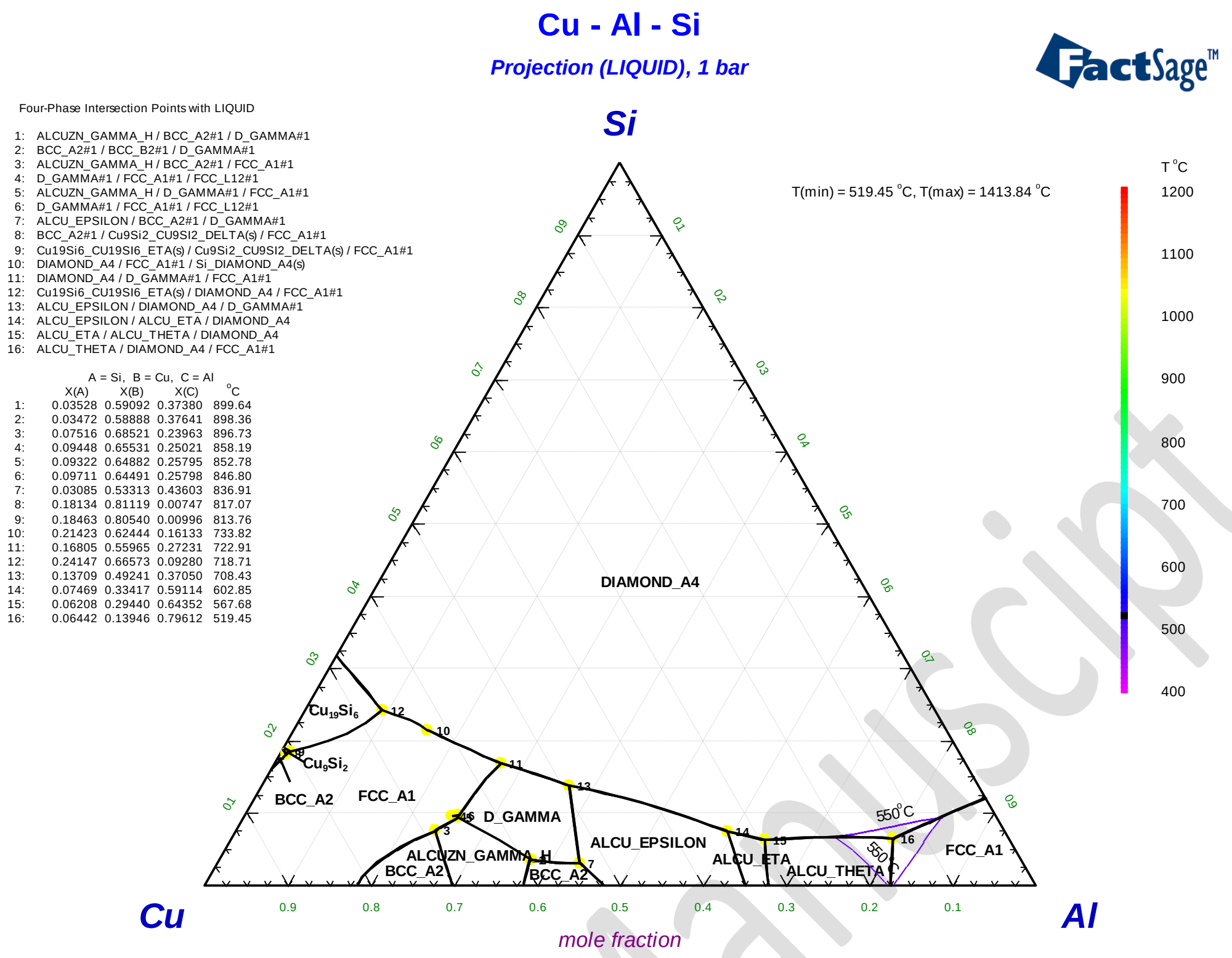

Figure 3-A Diagram with liquidus projections – SGTE without FCC2 phase (either FCC0 or FCC2 has to be selected)

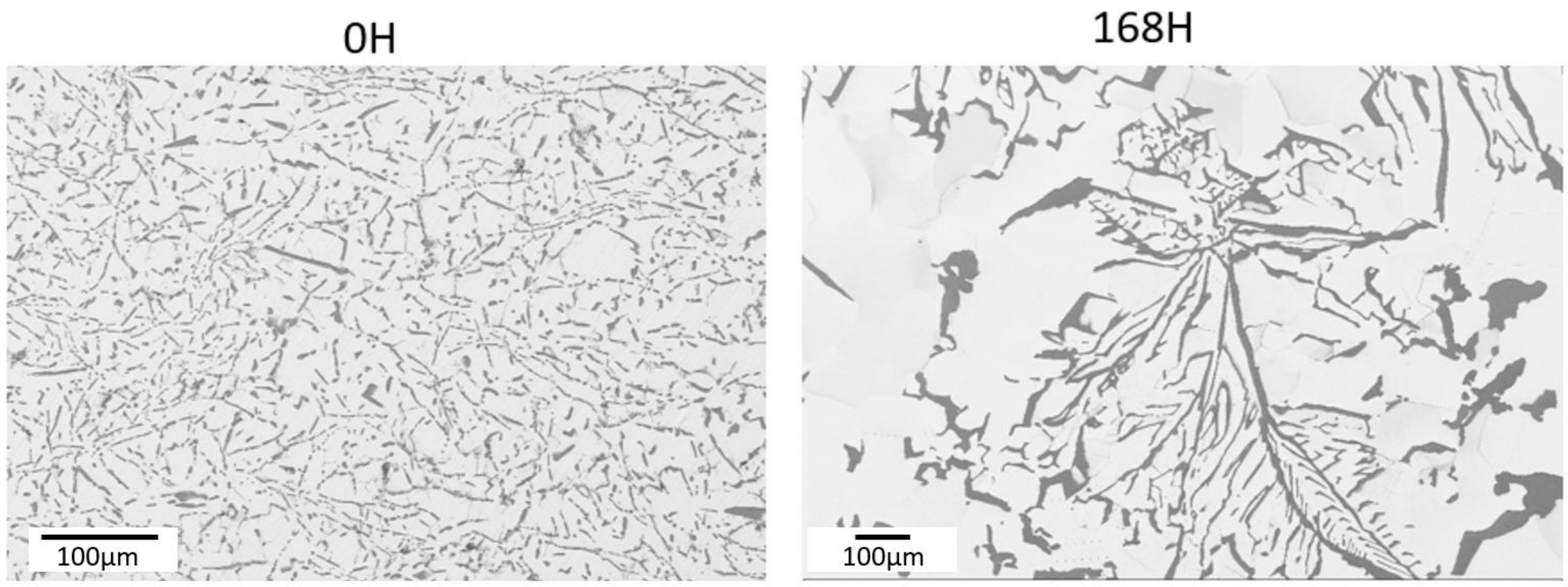

Figure 4-SEM microstructure of 84Cu-16Si after synthesis and after 168h at 840°C